\documentclass[aps,prc,twocolumn,showpacs,superscriptaddress]{revtex4-1}
\usepackage{hyperref}

\newcommand{\be}{\begin{equation}}
\newcommand{\ee}{  \end{equation}}
\newcommand{\ba}{\begin{eqnarray}}
\newcommand{\ea}{  \end{eqnarray}}
\newcommand{\bas}{\begin{eqnarray*}}
\newcommand{\eas}{  \end{eqnarray*}}

\begin{document}

\title{Effective Field Theory for Finite Systems with Spontaneously
Broken Symmetry}

\author{T. Papenbrock} 

\affiliation{Department of Physics and Astronomy, University of
  Tennessee, Knoxville, Tennessee 37996, USA}
\affiliation{Physics Division, Oak Ridge National Laboratory, Oak Ridge, Tennessee 37831, USA}

\author{H. A. Weidenm{\"u}ller}

\affiliation{Max-Planck-Institut f{\"u}r Kernphysik, D-69029
  Heidelberg, Germany}

\begin{abstract}
  We extend effective field theory to the case of spontaneous symmetry
  breaking in genuinely finite quantum systems such as small
  superfluid systems, molecules or atomic nuclei, and focus on
  deformed nuclei. In finite superfluids, symmetry arguments alone
  relate the spectra of systems with different particle numbers. For
  systems with non-spherical intrinsic ground states such as atomic
  nuclei or molecules, symmetry arguments alone yield the universal
  features of the low-lying excitations as vibrations that are the
  heads of rotational bands. The low-lying excitations in deformed
  nuclei differ from those in molecules because of symmetry
  properties caused by pairing.
\end{abstract}


\maketitle
\section{Introduction}

This paper addresses spontaneous symmetry breaking in non-relativistic
quantum systems of finite size. Strictly speaking, spontaneous
symmetry breaking can only occur in infinite systems. Then the ground
state exhibits a lesser degree of symmetry than the Hamiltonian
itself, i.e., the ground state is invariant under a symmetry group
${\cal H}$ that is a proper subgroup of the Hamiltonian's symmetry
group ${\cal G}$. The low-energy excitations are strongly constrained
by symmetry and given in terms of (weakly interacting) Nambu-Goldstone
modes. These can be calculated within an effective field theory (EFT)
that is solely based on the pattern of symmetry breaking. From a
technical point of view, the EFT is a nonlinear $\sigma$ model with
fields that parameterize the coset space ${\cal G}/{\cal
  H}$~\cite{weinberg1968,coleman1969,callan1969}. Examples for
spontaneous symmetry breaking are the breaking of spin-rotational
symmetry in a ferromagnet, the breaking of translational symmetry in a
crystal lattice, and the breaking of chiral symmetry in quantum
chromodynamics (QCD). In these examples, the Nambu-Goldstone modes are
magnons, phonons, and pions, respectively, and EFTs have been
developed for all these
cases~\cite{Gasser1984,Leutwyler1994,Leutwyler1996,roman1999,hofmann1999,Baer2004,kampfer2005},
see Refs.~\cite{weinbergbook,brauner2010} for reviews.

In finite systems, the ground state exhibits the full symmetry of the
Hamiltonian, and spontaneous symmetry breaking becomes evident in
symmetry-unrestricted mean-field
calculations~\cite{aberg1990,Nazarewicz1994,Yannouleas2007}. It is
then a major effort (and complication!) to restore the symmetry with
the help of projection techniques. The expressions ``obscured symmetry
breaking''~\cite{Koma1994} or ``emergent symmetry
breaking''~\cite{Yannouleas2007} (which we adopt here) have been
proposed for such systems. There are two distinct cases for which
emergent symmetry breaking plays a role. First, numerical simulations
of infinite physical systems are usually limited to a finite volume,
and it is then important to understand the finite-size corrections.
Some rigorous results are known in this
case~\cite{Horsch1988,Koma1994}, and finite-size corrections to
partition functions and thermodynamical observables have been worked
out within EFTs for simulations of QCD on finite
lattices~\cite{Leutwyler1987,Gasser1988}, and for spin
systems~\cite{Hasenfratz1993}. Genuinely finite systems constitute the
second and probably physically most interesting case. Prominent
examples are the emergence of superfluidity in trapped ultracold Bose
gases~\cite{matsumoto2002,Enomoto2006}, pairing in atomic nuclei (both
breaking a $U(1)$ phase symmetry), and non-spherical shapes of
molecules and atomic nuclei (both breaking $O(3)$ rotational symmetry
in the limit of infinite system size). Here, the techniques for
constructing EFTs for spontaneous symmetry breaking need to be
modified, and the interest is in spectra and transitions rather than
in thermodynamical observables.

The EFT for a finite system with emergent symmetry breaking is, of
course, related to the EFT for the corresponding infinite system with
spontaneous symmetry breaking. The symmetry must be realized
nonlinearly, and the Nambu-Goldstone fields parameterize the coset
space ${\cal G}/{\cal H}$~\cite{weinberg1968,coleman1969,callan1969}.
In the infinite system, the proper Nambu-Goldstone fields depend on
space and time and exhibit fluctuations of small amplitudes and long
wavelengths. A purely time-dependent (and spatially constant) mode is
forbidden because it would relate states of inequivalent Hilbert
spaces. In the finite system, however, this zero mode, i.e., the
spatially constant mode of the Nambu-Goldstone field, must be singled
out and treated separately. That mode undergoes large-amplitude
fluctuations and upon quantization restores the symmetry. In the
finite system, the small-amplitude fluctuations, i.e., the proper
Nambu-Goldstone modes with nontrivial temporal and spatial dependence,
must likewise be quantized. The theoretical implementation of this
program is not trivial and is demonstrated for two interesting and
important cases. We first consider as an example the emergent breaking
of a $U(1)$ phase symmetry in finite superfluids such as ultracold
bosonic atom gases or atomic nuclei. In this case, the proper
Nambu-Goldstone modes and the global phase rotations do not couple in
leading order, and both modes have the same energy scale. This
facilitates the description. Second, we consider the emergent breaking
of $SO(3)$ symmetry to its $SO(2)$ subgroup. This case describes the
low-energy physics of nonspherical objects with axial symmetry such as
linear molecules and deformed atomic nuclei. The case is more
complicated and interesting due to the interactions between global
rotations and proper Nambu-Goldstone modes, and the energy scale of
the rotational mode differs from the energy scale of the proper
Nambu-Goldstone modes. Our detailed presentation of these two problems
makes clear how to develop EFTs for systems with emergent symmetry
breaking in general. 

In this paper we construct EFTs for finite systems with emergent
symmetry breaking, and we focus particularly on deformed atomic
nuclei. Such nuclei are traditionally described by generalized
collective
models~\cite{bohr_1952,bohrmottelson_1953,eisenberg,bmbook,rowe} or
the interacting boson model~\cite{arima1975,iachello}.  For more
microscopic approaches to collective motion, we refer the reader to
Refs.~\cite{aberg1990,frauendorf2001,niksic2011}.  For deformed
nuclei, the presented EFT generalizes the simpler construction of an
effective theory proposed recently~\cite{papenbrock2011}. Based on
symmetry principles alone, our model-independent approach re-derives
some of the well-known results for collective nuclear models.  We
expect that extensions of the EFT approach could be useful in
addressing well-known and long-standing limitations of the collective
models such as, e.g., the significant overprediction of
electromagnetic interband transitions, see
Refs.~\cite{garrett2001,rowe} for recent reviews of this problem.

Our procedure is patterned after the case of the infinite
ferromagnet~\cite{Leutwyler1994,roman1999, kampfer2005}.  We
generalize the expression for unitary transformations in the coset
space by including purely time-dependent variables. These account for
the dynamics of the finite system. The resulting generators define the
Nambu-Goldstone modes as well as the zero modes as classical fields.
From these we construct the building blocks of the effective
Lagrangian $L$ using arguments of invariance and energy scaling.
Quantization of the Hamiltonian obtained by a Legendre transformation
of $L$ then determines the spectrum.

\section{Emergent breaking of U(1) phase symmetry}
\label{emer}

Superfluids can be viewed as breaking $U(1)$ phase symmetry. Examples
are infinite Bose-Einstein condensates (BEC) or the paired states of a
BCS superconductor. In their mean-field description, these systems
exhibit a coherent phase at the expense of a well-defined particle
number. In finite systems the particle number is, of course, a good
quantum number. While the results we derive in this Section are well
know (see, e.g. Ref.~\cite{wen}), their derivation exhibits novel
aspects and paves the way for the description of deformed nuclei
within an EFT.

In finite superfluids the low-lying excitations are governed by two
energy scales. These are the chemical potential $\mu$ (the energy
needed to add a single boson to the system), and the energy $\Omega$
of long wave-length excitations. These scales are different for
noninteracting and for interacting systems. We consider harmonically
trapped bosons as an example. For noninteracting bosons, the proper
thermodynamic limit is defined~\cite{dalfovo1999} by keeping the
product of particle number $N$ and the third power of the trap
frequency constant. That frequency, in turn, defines $\mu$ for the
condensate. Hence $\mu \sim \Omega \sim N^{-1/3}$ scale similarly. As
$N \to \infty$, the ground states of systems with different particle
numbers $N$ become quasi degenerate, superpositions of such states
(i.e., states with a constant phase and undefined $N$) describe the
superfluid, and the $U(1)$ phase symmetry is broken in the
thermodynamic limit.  For finite uniform systems, both energies scale
as $\mu\sim\Omega\sim N^{-2/3}$~\cite{huang1987}, and the arguments
apply likewise.

For interacting Bose gases of volume $V$, the chemical potential $\mu$
typically approaches in the thermodynamic limit the nonzero value $\mu
\sim g N/V$ where $g$ measures the strength of the
interaction~\cite{dalfovo1999}. Similarly, for BCS superconductors,
adding a pair roughly costs an energy of the order of the Fermi
energy. The latter becomes constant in the thermodynamic limit. In
both cases the breaking of $U(1)$ symmetry can be understood in the
framework of the grand canonical ensemble. The system is coupled to a
particle reservoir with external chemical potential $\mu_{\rm ext}$,
and the Hamiltonian $H-\mu_{\rm ext}N$ is minimized. Adjusting
$\mu_{\rm ext}$ such that $\mu_{\rm ext}\approx\mu$ for $N \gg 1$
introduces a quasi degeneracy between states of different particle
numbers, and a superposition of these states then breaks $U(1)$
symmetry. However, the canonical and the grand canonical ensemble
yield different results. For nonextensive quantities the differences
decrease as $N^{-1/2}$ for $N \to \infty$. It is only within this
uncertainty that an isolated finite system can be viewed as equivalent
to a finite system coupled to a particle reservoir. Technically, the
introduction of a finite external chemical potential $\mu_{\rm ext}$
breaks time-reversal invariance, and the resulting effective theory
differs from the case $\mu_{\rm ext} = 0$. As we will see, the
latter can be recovered from the former by simply setting $\mu_{\rm
  ext} = 0$ in the leading-order equations we derive below. For
nonzero $\mu_{\rm ext}$, the low-energy scales of interest are $\mu
-\mu_{\rm ext}$ and $\Omega$, and we assume that both are similar in
size.

In the case of a broken $U(1)$ symmetry we have ${\cal G} = U(1)$ and
${\cal H} = 1$. The Nambu-Goldstone fields parameterize the coset
${\cal G}/{\cal H}\sim {\cal G}$, which is the group itself. The
fields induce local phase transformations, and in a finite system the
relevant operator is
\be
U(\alpha,\beta) = e^{i\alpha(t)}e^{iV^{1/2}\beta(t,\vec{x})} \ . 
\label{uni}
\ee
Here, $\beta$ is the Nambu-Goldstone field, the angle $\alpha$ is the
zero mode that needs to be singled out, and $V$ is the volume. For a
proper Nambu-Goldstone field we have $\int_V {\rm d}^3x \beta = 0$.
Following Refs.~\cite{weinbergbook,brauner2010}, we build the
invariants of our theory from the derivatives ($\nu = x, y, z$) 
\ba
-i U^{-1} \partial_t U &=& \dot\alpha +V^{1/2} \dot\beta \ , \\ 
-i U^{-1} \partial_\nu U &=& V^{1/2} \partial_\nu \beta 
\ea
which are in the Lie algebra of ${\cal G}$. Here the dot denotes the
time derivative. Under a global phase transformation with angle
$\gamma$, the operator $U$ becomes $e^{i\gamma}U(\alpha,\beta) =
U(\alpha+\gamma,\beta)$. Thus, $\dot\alpha$, $\dot\beta$, and
$\partial_\nu \beta$ are invariant under global phase transformations.
Note also that $\beta$ is a truly ``intrinsic'' degree of freedom
because it is unaffected by a global phase transformation.

We use these invariants to construct the leading-order terms in the
effective Lagrangian, taking account of energy scales. The scale
associated with the $\alpha$ degree of freedom is $\mu - \mu_{\rm
    ext}$, that associated with $\beta$ is
$\Omega$. Assuming also invariance under rotations, we have the
invariants
\ba
L_0 &\equiv&  C_1\mu_{\rm ext}\int\limits_V {\rm d}^3x \dot\alpha
=  C_1V\mu_{\rm ext}{\dot\alpha} \ , \nonumber \\
L_1 &\equiv& {C_1\over 2} \int\limits_V {\rm d}^3x \dot\alpha^2 =
{C_1V\over 2}{\dot\alpha}^2 \ , \nonumber\\ 
L_2 &\equiv& {C_2 \over 2} \int\limits_V {\rm d}^3x \dot\beta^2 \ , \ 
L_3 \equiv {D \over 2} \int\limits_V {\rm d}^3x \left(\nabla \beta
\right)^2 \ .
\ea
Here, $C_1$, $C_2$, and $D$
are constants that have to be determined from low-energy data. The
Lagrangian is
\be
\label{eff} 
L = {C_1 V \over 2}{\dot\alpha}^2 + C_1 V \mu_{\rm ext} \dot\alpha +
\int\limits_V {\rm d}^3x \left( {C_2\over 2} \dot\beta^2 -{D\over 2}
\left(\nabla\beta\right)^2 \right) \ .
\ee
The conserved quantity corresponding to invariance under global phase
transformations is the particle number
\be
\label{nparticle}
N\equiv p_\alpha \equiv {\partial L\over \partial\dot\alpha} \ . 
\ee
This conserved quantity can be derived via the Noether theorem, and
$p_\alpha$ is the canonical momentum of $\alpha$.
 
We expand $\beta(t,\vec{x}) = \sum_j \beta_j \phi_j(\vec{x})$ in a set
of orthonormal complex functions $\phi_j(\vec{x})$, $j=1, 2, \ldots$
with $\int_V {\rm d}^3 x \phi_j(\vec{x}) = 0$ (absence of zero modes
for the Nambu-Goldstone field). As an example, we choose the
eigenfunctions of a free particle in a spherical cavity of volume $V$
with von Neumann boundary conditions. The Lagrangian becomes
\be
L = {C_1 V\over 2} {\dot\alpha}^2 + C_1 V \mu_{\rm ext}\dot\alpha +
\sum_{j>0}\left({C_2\over 2}{\dot\beta}_j^2 - {D k_j^2\over 2}\beta_j^2
\right) \ . 
\ee
Here, $k_j^2$ denotes the squared momentum of the spherical wave
$\phi_j(\vec{x})$. A Legendre transformation with $p_{\beta j} \equiv
\partial L / \partial {\dot\beta}_j$ and Eq.~(\ref{nparticle}) yield
the Hamiltonian
\be
H = {\left(p_\alpha - C_1V\mu_{\rm ext}\right)^2\over 2C_1V} + \sum_{j>0}
\left({p_{\beta j}^2\over 2C_2}  + {D k_j^2 \over 2}\beta_j^2 \right) \ . 
\ee
We quantize $H$ by putting $p_\alpha = -i\partial_\alpha$, and
$p_{\beta j} = -i \partial_{\beta_j}$. Then $p_\alpha e^{iN\alpha} = N
e^{iN\alpha}$. The intrinsic degrees of freedom $\beta_j$ yield
harmonic-oscillator spectra with energies
\be
\label{evib}
\omega_j \equiv k_j \left({D\over C_2}\right)^{1/2} \ .
\ee 
In the long-wave-length limit we have $k_j \sim V^{-1/3}$, and the
measurement of the low-energy collective excitations of the superfluid
determines the ratio $D/C_2$. The amplitude of the Nambu-Goldstone
modes (i.e., the oscillator length) is
\be
\label{osclen}
l_j\equiv \left(C_2Dk_j^2\right)^{-1/4} \ ,
\ee
and this dimensionless quantity is assumed to be small, $l_j \ll 1$.

For fixed particle number $N$, the $\alpha$-dependent part of the
Hamiltonian yields the energy $E_N \equiv \left(N-C_1V\mu_{\rm ext}
\right)^2 / (2C_1V)$. For $N \gg 1$ the energy difference of a system
with $N + 1$ and one with $N$ particles is
\be
\label{chempot}
E_{N+1}-E_N \approx {N \over C_1 V} - \mu_{\rm ext} \ . 
\ee
We note that $E_{N+1}-E_N = \mu-\mu_{\rm ext}$ on physical
  grounds.  The emergent breaking of $U(1)$ requires
  $E_{N+1}-E_N\approx 0$, and Eq.~(\ref{chempot}) relates the constant
  $C_1$ to the chemical potential and to the density of the
  system. We see that the chemical
potential and the frequencies and amplitudes of the quantized
collective vibrations determine the low-energy constants of the EFT.
In addition to the collective vibrations with frequencies~(\ref{evib})
one finds approximately equidistant levels with spacing
  $\mu - \mu_{\rm ext}$ belonging to superfluids with different
particle numbers. In superfluid atomic nuclei, these harmonic
excitations belonging to different numbers of paired nucleons are
known as pairing vibrations~\cite{Bes1966}. In summary we have shown
that in finite superfluids, the spectra of systems with different
particle numbers are related to each other, and this is a
model-independent prediction of the EFT.

We turn to higher-order corrections
and establish our power counting.  The energy scales used in the
construction of $L$ in Eq.~(\ref{eff}) are (i) the scale $\mu -
\mu_{\rm ext}$ associated with a change in particle number (we have
assumed $\dot\alpha \sim \mu-\mu_{\rm ext}$) and (ii) the scale
$\Omega$ of the collective vibrations. We have assumed that both
scales are of similar size, $\mu - \mu_{\rm ext} \sim \Omega$.
Moreover, in the low-energy domain the amplitudes of the
Nambu-Goldstone modes $\beta_j$ given by Eq.~(\ref{osclen}) obey $l_j
\ll 1$. Thus, we have
\ba
\label{scalingu1}
\beta_j \sim l_j \ , \ {\dot\beta}_j \sim \Omega l_j \ , C_2 \sim
\Omega^{-1} l_j^{-2} \ , \ p_{\beta j} \sim l_j^{-1} \ .
\ea
In the construction of terms of higher order, we consider only powers
of single derivatives because higher-order derivatives of the
Nambu-Goldstone field $\beta$ and of the zero mode $\alpha$ can be
eliminated via perturbative field
redefinitions~\cite{damour1991,grosseknetter1994}. The EFT has a
breakdown scale $\Lambda$ with $\Lambda \gg \Omega, |\mu_{\rm ext}
-\mu|$. At this scale, $\dot\alpha$ (and $\dot\beta$, $\partial_\nu
\beta$, $\beta$) are a factor $\sqrt{\Lambda/|\mu_{\rm ext}-\mu|}$
(and $\sqrt{\Lambda/\Omega}$, respectively) larger than in the
low-energy domain. At the breakdown scale, a Lagrangian term of the
form $C_{pqr}{\dot\alpha}^p{\dot\beta}^q \beta^r$ must scale as
$\Lambda$ when written in terms of such velocities and fields. This
determines the scaling of $C_{pqr}$.  When evaluated for velocities
and fields in the low-energy domain, that term yields a contribution
to $L$ of order $\Lambda (|\mu_{\rm ext} - \mu| / \Lambda)^{p/2}
(\Omega/\Lambda)^{q/2+r/2}$. Similar considerations apply for terms
containing powers of the spatial derivatives $\partial_\nu\beta$. For
energies below the breakdown scale, all these higher-order corrections
are perturbatively small. Thus, our procedure yields a perturbative
expansion in powers of $|\mu_{\rm ext}-\mu|/\Lambda$ and
$\Omega/\Lambda$.

\section{Emergent breaking of rotational symmetry}

For finite objects with an axially symmetric ground-state deformation,
the emergent symmetry breaking is from the rotation group to the
subgroup of axial symmetry. It is, therefore, useful to recall the
breaking of spin-rotational symmetry $O(3)$ to $O(2)$ in infinite
ferromagnets~\cite{Leutwyler1994,roman1999, Baer2004}. In the ground
state, all spins point in the same direction, violating the
spin-rotational symmetry of the Hamiltonian. Ground states with
macroscopically different spin directions have zero overlap and define
inequivalent Hilbert spaces. As a consequence, the low-lying spectrum
of the ferromagnet is dominated by Nambu-Goldstone modes, i.e., spin
waves of long wave length that locally induce small rotations of the
aligned spins. For a ferromagnet of finite size~\cite{Hasenfratz1993},
the formerly inequivalent Hilbert spaces are connected by
non-vanishing tunneling matrix elements, the ground states belonging
to different spin directions have nonzero overlap, and there exist
states that are superpositions of these ground states. Such states
are, for instance, the Wigner $D$-functions for rotational motion.
Physically that implies that the ferromagnet may rotate about an axis
perpendicular to the direction of the aligned spins. An analogous
situation occurs in linear molecules and in axially symmetric deformed
even-even nuclei. In the limit of infinitely large size, the low-lying
parts of the spectra of these systems would be determined by
Nambu-Goldstone modes. But finite linear molecules or finite axially
symmetric nuclei may rotate about an axis perpendicular to the
symmetry axis. In all three cases, rotational motion occurs as a
consequence of emergent symmetry breaking due to the finite size of
the system. Rotational motion is, therefore, distinctly different from
the Nambu-Goldstone modes. It plays the same role as the zero mode
$\alpha$ in Eq.~(\ref{uni}) for the emergent breaking of $U(1)$ phase
symmetry.

For a deformed nucleus with $A \gg 1$ nucleons, we can quantify these
statements. The linear extension of the system is $\propto A^{1/3}$.
The moment of inertia is proportional to mass $\times$ (length)$^2
\propto A^{5/3}$. With increasing $A$, the frequency of rotational
motion tends to zero like $A^{- 5/3}$, faster than the wave number
$\propto A^{- 1/3}$ of the massless modes (analogues of the
Nambu-Goldstone modes in the infinite system). Hence, there exists a
regime of $A$ values where the scale $\xi$ for rotational motion is
small compared to the scale $\omega$ for vibrational motion and where
$\omega$ in turn is small compared to the breakdown scale $\Lambda$ of
the EFT defined by pair-breaking excitations~\cite{dean2003}. In the
regime $\xi \ll \omega \ll \Lambda$, it is meaningful to consider
rotational motion as a small (in energy) correction to the Goldstone
theorem. That is the regime we study here. For instance, in rare-earth
nuclei we have $\xi \approx 80$~keV from the lowest energy spacing in
a rotational band, and $\omega \approx 1$~MeV from the lowest
``vibrational'' band head, while $\Lambda \approx 2-3$~MeV is the cost
of pair breaking. The condition $\omega \ll \Lambda$ is met only
marginally. Corresponding considerations apply to linear molecules
such as CO$_2$. Here the scale of rotational energies is $\xi
\approx$~1~cm$^{-1}$, that of vibrational energies is $\omega
\approx$~500~cm$^{-1}$, while the breakdown scale
$\Lambda\approx$~10,000~cm$^{-1}$ is defined by electronic
excitations. The conditions $\xi \ll \omega \ll \Lambda$ are very well
fulfilled.

Axially symmetric deformed even-even nuclei consist of nucleons, and
linear molecules consist of nuclei. The locations of these
constituents have body-fixed spherical coordinates $r, \theta, \phi$.
Vibrations of the nucleus/molecule about an axis perpendicular to the
symmetry axis act locally on the constituents. The resulting
dislocations are assumed to have small amplitude. We expect that
Nambu-Goldstone modes related to the coordinate $r$ have higher
frequencies than those due to $\theta$ and $\phi$. We, therefore,
confine attention to the latter variables although our approach can be
straightforwardly generalized. In addition to these small-amplitude
vibrations we also consider global rotations of the entire
nucleus/molecule.

Our effective theory is universal and applies both to linear molecules
and to deformed nuclei. Apart from the magnitude of the low-energy
constants, the key difference between both systems is in the symmetry
properties of the deformed ground--state wave functions as these
define the symmetry properties of the admissible low--lying
Nambu-Goldstone modes~\cite{weinbergbook}. We assume that molecular
and nuclear ground-state wave functions are axially symmetric about
the body-fixed $z'$-axis, invariant under time-reversal, and have
positive parity. As a consequence of pairing (superfluidity),
even-even nuclei differ from linear molecules in that their intrinsic
ground states are also invariant under rotations by $\pi$ about any
axis perpendicular to the symmetry axis, i.e., possess positive ${\cal
  R}$ parity~\cite{bohr1958,bmbook}. Hence, low-energetic intrinsic
excitations in nuclei must also have positive ${\cal R}$ parity.

\subsection{Dynamical Variables and Power Counting}

As done for the case of the
ferromagnet~\cite{Leutwyler1994,roman1999,kampfer2005} and in
Sect.~\ref{emer}, we consider the Nambu-Goldstone modes as classical
fields that are later quantized.  We prefer to write these fields in
the space-fixed (rather than the body-fixed) coordinate system because
here the commutation relations of the three generators $J_x, J_y, J_z$
of infinitesimal rotations about the space-fixed $x, y, z$ axes,
respectively, are of standard form. The molecular/nuclear ground state
is invariant under $SO(2)$ rotations about the body-fixed $z'$-axis
while $SO(3)$ symmetry is broken by the deformation. Therefore, the
Nambu-Goldstone modes lie in the coset space $SO(3) /
SO(2)$~\cite{coleman1969,callan1969,brauner2010}. The modes depend on
the angles $\theta, \phi$ defined above and on time $t$, and are
generated by a unitary transformation $U$. As in Eq.~(\ref{uni}) we
parameterize the matrix $U$ in product form,
\ba
\label{1}
U &=& g(\alpha, \beta) u(x,y) \ , \nonumber\\
g(\alpha, \beta) &=& \exp \left\{ -i \alpha(t) J_z \right\}
\exp\left\{ -i \beta(t) J_y \right\} \ , \nonumber\\
u(x,y) &=& \exp\left\{-i x(\theta, \phi, t) J_x
-i y(\theta, \phi, t) J_y \right\} \ .
\ea
The purely time-dependent variables $\alpha(t)$ and $\beta(t)$ are the
zero modes. They describe global rotations of the finite system and
are factored out~\cite{Leutwyler1987}. They are not Nambu-Goldstone
modes but upon quantization generate rotational
bands~\cite{chandrasekharan2008,papenbrock2011}. The fields $x(\theta,
\phi, t)$ and $y(\theta, \phi, t)$ with $|x|, |y| \ll 1$ generate
small-amplitude vibrations of the constituents. These depend
non-trivially on $\theta$ and $\phi$ so that
\be
\label{2}
\int {\rm d} \Omega \ x(\theta, \phi, t) = 0 = \int {\rm d} \Omega \
y(\theta, \phi, t) \ .
\ee
Here ${\rm d} \Omega$ is the surface element of the three-dimensional
unit sphere. In an infinite system $x(\theta, \phi, t)$ and $y(\theta,
\phi, t)$ would be genuine Nambu-Goldstone modes.  Eqs.~(\ref{1}) and
(\ref{2}) define the dynamical variables of the system. Eq.~(\ref{1})
may look like a rather special ansatz but actually follows from the
most general form of $U$.

Further progress hinges on the identification of the energy scales
$\xi$, $\omega$, and $\Lambda$ defined above. The ranges of the
variables $\alpha$ and $\beta$ being of order unity, the ratios
$\dot{\alpha} / \alpha$ and $\dot{\beta} / \beta$ are governed by the
energy scale $\xi$ of rotational motion. We have $|x|, |y| \ll 1$,
indicating that the amplitudes of the Nambu-Goldstone fields are small.
Then $|\dot{x}|\sim \omega |x|$ and $|\dot{y}| \sim \omega|y|$.  We
are going to show that power counting based upon the inequalities $\xi
\ll \omega \ll \Lambda$ together with the symmetry requirements
formulated above uniquely determine the leading-order part of the
Hamiltonian, except for a small number of constants that have to be
determined by fits to data.

Our EFT is characterized by two breakdown scales. The first scale is
set by $\Lambda \gg \omega$ and marks the appearance of neglected
degrees of freedom. In deformed nuclei these are single-particle
degrees of freedom or pair-breaking effects and in linear molecules,
electronic excitations. The second scale is related to large-amplitude
excitations of the Nambu-Goldstone fields. That scale is reached when
these excitations are so large that they practically restore the
spherical symmetry of the intrinsically deformed object, or when the
energy due to the zero-mode velocities $\dot\alpha$ and $\dot\beta$
reaches the vibrational scale $\omega$. This second scale is set by
$\omega^2/\xi$. For well-deformed nuclei, that scale considerably
exceeds $\Lambda$, while both scales are similar in size for linear
molecules. We now discuss both breakdown scales separately.

At energies below $\Lambda$ the neglected degrees of freedom cause the
appearance of higher-order terms in the effective Lagrangian of the
EFT. Such terms involve powers of the leading-order fields and
velocities and, possibly, also higher derivatives. The latter can be
eliminated via perturbative field
redefinitions~\cite{damour1991,grosseknetter1994} and are not
considered here. At the breakdown scale (where the amplitudes and
velocities of the Nambu-Goldstone fields are a factor
$(\Lambda/\omega)^{1/2}$ larger than in the low-energy domain), a term
in the effective Lagrangian with $\tau$ velocities and $n$ powers of
$x$ or $y$ is of order $\Lambda$. At the low-energy scale $\omega$
that term yields a contribution of order $\omega (\omega /
\Lambda)^{(n+\tau)/2-1}$. Terms like that give rise to small corrections
and, at each order, are finite in number. Similar considerations apply
to the spatial derivatives.

The scale for the breaking of emergent symmetry is reached when the
amplitudes, velocities and higher derivatives are a factor $(\omega /
\xi)^{1/2}$ larger than in the low-energy domain. The arguments of the
previous paragraph can essentially be repeated by replacing the scale
$\Lambda$ by $\omega^2/\xi$, and the ratio $\Lambda / \omega$ by
$\omega / \xi$. At the breakdown scale the contribution of a term in
the effective Lagrangian which contains $\sigma$ velocities is of
order $\omega$. At the low-energy scale $\xi$ that term scales as $\xi
(\xi / \omega)^{\sigma/2-1}$.

These arguments establish our power counting. As a result, our EFT
provides an expansion in the two small parameters $\omega / \Lambda$
and $\xi / \omega$, and there is a finite number of terms for each
power of these parameters.

\subsection{Effective Lagrangian} 

The effective Lagrangian is built from invariants. These are
constructed from elements $a_\mu^x, a_\mu^y, a_\mu^z$ defined by
\ba
\label{3}
U^{-1} i \partial_\mu U &=& a_\mu^x J_x + a_\mu^y J_y + a_\mu^z J_z \ .
\ea
The symbol $\partial_\mu$ with $\mu = 1, 2, 3$ stands for the partial
derivatives with respect to the angles $\theta, \phi$ and time $t$.
Explicit expressions for these elements are obtained from
Eqs.~(\ref{1}) and (\ref{3}) in terms of a series expansion in powers
of $x$, $y$, and their partial derivatives where only leading-order
terms are kept. We use the Baker--Campbell--Hausdorff
expansion and obtain
\ba
a^x_t &=& \dot{x} +{y \over 6} (x \dot{y} - y \dot{x}) -\dot{\alpha}
\sin \beta - y \dot{\alpha} \cos \beta + \ldots\ , \nonumber \\
a^y_t &=& \dot{y} -{x \over 6} (x \dot{y} - y \dot{x}) + \dot{\beta}
+ x \dot{\alpha} \cos \beta + \ldots \ , \\
\label{3a}
a^z_t &=& -{1 \over 2} (x \dot{y} - y \dot{x}) + \dot{\alpha} \cos
\beta - y \dot{\alpha} \sin \beta - x \dot{\beta} + \ldots \ , \nonumber
\ea
and 
\ba
a^x_\nu &=& \partial_\nu{x} +{y \over 6} (x \partial_\nu{y} - y \partial_\nu{x}) + \ldots\ , \nonumber \\
a^y_\nu &=& \partial_\nu{y} -{x \over 6} (x \partial_\nu{y} - y \partial_\nu{x}) + \ldots \ , \\
\label{3b}
a^z_\nu &=& -{1 \over 2} (x \partial_\nu{y} - y \partial_\nu{x}) + \ldots \ , \nonumber
\ea
To define the invariants we calculate the changes induced on the
variables $\alpha, \beta, x, y$ by infinitesimal rotations $r$ of $U$
about angles $\delta \chi_k$ around the space-fixed $k = x, y, z$
axes.  With $h(\gamma) = \exp \{ - i \gamma J_z \}$ we have from
Eq.~(\ref{1}) that
\be
\label{rot1}
r g(\alpha, \beta) = g(\alpha', \beta') h(\gamma') \ .
\ee
Here, the primed angles depend on the angles of the rotation $r$ and
the angles $\alpha$, $\beta$.  The right-hand side of Eq. (\ref{rot1})
has the form of a general rotation with Euler angles ($\alpha',
\beta',\gamma'$), with explicit expressions given in
Ref.~\cite{papenbrock2011}. Thus,
\ba
r U 
&=&g(\alpha', \beta') h(\gamma') u \nonumber\\
&=&g(\alpha', \beta') \ [ h(\gamma') u h^\dag(\gamma') ]\ h(\gamma') \ . 
\ea
As a result we find that the angles $\alpha$ and $\beta$ transform
nonlinearly as the azimuthal and polar angle of the two-sphere,
respectively, while $x$ and $y$ transform linearly as the $x$ and $y$
components of a vector under rotations around the $z$ axis. In other
words, under a rotation, the nucleus as a whole changes its
orientation and undergoes a rotation around its symmetry axis. This
transformation behavior under rotations confirms that $\alpha$ and
$\beta$ describe the global orientation of the axially symmetric
nucleus, while $x$ and $y$ are ``intrinsic'' degrees of freedom.  Thus
any combination of $x$ and $y$ that formally exhibits axial symmetry
is indeed fully invariant under rotations. For example, $x^2 + y^2$ is
invariant under rotations, and the four quantities $x, y, \dot{x},
\dot{y}$ are transformed into linear combinations of $x', y',
\dot{x}', \dot{y}'$. These transformation properties are
characteristically different from the ones for an infinite system
where in Eq.~(\ref{1}) we would have $g(\alpha, \beta) = 1$.

Time-reversal invariance requires that invariants involving time
derivatives must contain even powers of $a^x_t, a^y_t, a^z_t$. The
lowest-order invariants obtained from Eqs.~(\ref{3a}) are
\ba
\label{4}
{\cal L}_{1a} &=& \dot{\beta}^2 + \dot{\alpha}^2 \sin^2 \beta \ ,
\nonumber \\
{\cal L}_{1b} &=& \dot{x}^2 + \dot{y}^2 + 2(x \dot{y} - y \dot{x})
\dot{\alpha} \cos \beta \ , \nonumber \\
{\cal L}_{1c} &=& (x \dot{y} - y \dot{x})^2 \ , \nonumber \\
{\cal L}_{1d} &=& (x^2 +y^2)[\dot{x}^2 + \dot{y}^2 + 2(x \dot{y} - y
\dot{x}) \dot{\alpha} \cos \beta] \ .
\ea
We note that the invariant ${\cal L}_{1a}$ is essentially the
Lagrangian of a rotor, and that the Lagrangian density ${\cal L}_{1b}$
couples global rotations to the Nambu-Goldstone modes.  The invariant
${\cal L}_{1 c}$ is related to the angular momentum of the
Nambu-Goldstone modes, see Eq.~(\ref{K}).  The invariant ${\cal
  L}_{1d}$ is obtained by multiplying ${\cal L}_{1b}$ with the
invariant $(x^2 + y^2)$ and is of the same order as ${\cal L}_{1c}$.
As for the invariants constructed from $a^x_\nu$, $a^y_\nu$ $a^z_\nu$
with $\nu = \theta$ or $\nu = \phi$, we use that for fixed $\nu$ and
$\nu'$, the forms $(a^x_\nu)^2 + (a^y_\nu)^2$, $a^z_\nu$ and $a^z_\nu
a^z_{\nu'}$ are invariant. Admissible linear combinations of these
expressions are defined by the requirement of axial symmetry.
Suppressing terms of higher order than $x^4$ and multiplying with the
additional invariant $(x^2 + y^2)$, we find the invariants
\ba
\label{5}
{\cal L}_{2a} &=& (\vec{\bf L} x)^2 + (\vec{\bf L} y)^2 \ , \nonumber \\
{\cal L}_{2b}&=& ({\bf L}_z x)^2 + ({\bf L}_z y)^2 \ , \nonumber \\
{\cal L}_{2c} &=& (x \vec{\bf L} y - y \vec{\bf L} x)^2 \ , \nonumber \\
{\cal L}_{2d} &=& (x^2 + y^2) \left( (\vec{\bf L} x)^2 + (\vec{\bf L}
y)^2 \right) \ . 
\ea
Here $\vec{\bf L}$ (${\bf L}_z)$ is the vector (the $z$-component)
operator of orbital angular momentum, respectively, written in terms
of $\theta$ and $\phi$~\cite{varshalovich1988}. The occurrence of the
term ${\cal L}_{2 b}$ reflects the fact that we impose only axial
rather than rotational symmetry on ${\cal L}$. The Lagrangian $L$ is
given by
\ba
\label{6}
L &=& L_1 + L_2 \nonumber\\
&=& \sum_{i = a, b, c, d} \int {\rm d} \Omega \ \bigg(
\frac{C_i}{2} {\cal L}_{1 i} - \frac{D_i}{2} {\cal L}_{2 i} \bigg) \ .
\ea
Here $C_i$ and $D_i$ with $i = a, b, c, d$ are constants that are
determined by low-energy data.

We expand the real variable $x$ as
\ba
\label{7}
x = \sum_{\lambda = 2}^\infty \sum_{\mu = -\lambda}^\lambda
x_{\lambda \mu} Z_{\lambda \mu}
\ea
and correspondingly for $y$, $\dot{x}$, $\dot{y}$. Aside from
normalization constants, the real orthonormal functions $Z_{\lambda
  \mu}$ are equal to the real part (for $\mu\ge 0$) and imaginary part
(for $\mu<0$) of the spherical harmonics $Y_{\lambda \mu}$. The
coefficients $x_{\lambda\mu}$ are real. Terms with $\lambda = 0$ and
$\lambda = 1$ are excluded since $\lambda = 0$ violates Eq.~(\ref{2})
and describes global rotations while $\lambda = 1$ describes
translations in space~\cite{bmbook}. We insert the
expansions~(\ref{7}) into Eq.~(\ref{6}) and use the resulting
expression for $L$ to define the real canonical momenta
\ba
\label{8}
p_\beta &=& \frac{\partial L}{\partial \dot{\beta}} \ , \ p_\alpha
= \frac{\partial L}{\partial \dot{\alpha}} \ , \nonumber\\
p^x_{\lambda \mu} &=& \frac{\partial L}{\partial \dot{x}_{\lambda \mu}} \ , \
p^y_{\lambda \mu} = \frac{\partial L}{\partial \dot{y}_{\lambda \mu}} \ .
\ea
From Noether's theorem we obtain explicit expressions for the three
components $I_x, I_y, I_z$ of angular momentum. These are
\ba
I_x &=& - p_\beta \sin \alpha - p_\alpha \cot \beta \cos \alpha +
{\cos \alpha \over \sin \beta} \ K \ , \\
I_y &=& p_\beta \cos \alpha - p_\alpha \cot \beta \sin \alpha + {\sin
\alpha \over \sin \beta} \ K \ , \\
I_z &=& p_\alpha \ .
\ea
Here
\ba
\label{K}
K = \int {\rm d} \Omega \ (x p_y - y p_x ) 
\ea
is the angular momentum of the two-dimensional oscillators that
describe the intrinsic vibrations. The square of the total angular
momentum is
\be
\label{9}
I^2 = p_\beta^2 +{1 \over \sin^2\beta} \left(p_\alpha^2 -2Kp_
\alpha \cos \beta + K^2 \right) \ . 
\ee
The terms in Eqs.~(\ref{9}) obtained by putting $K = 0$ can be shown
to be equal to the square of the total angular momentum of the rigid
rotor.

\subsection{Effective Hamiltonian and Quantization} 

We use Eqs.~(\ref{8}) and the standard Legendre transformation to
transform the effective Lagrangian $L$ into the effective Hamiltonian
$H$. The scales of the coefficients $C_i$ and $D_i$ (and the terms
that are kept in $H$) are determined by assuming $C_a {\cal L}_{1a}
\sim \xi$, $C_b {\cal L}_{1b} \sim \omega$, $C_c {\cal L}_{1c}, C_d
{\cal L}_{1d} \sim |x|^2 \omega$, $p_\beta, p_\alpha \sim 1$, and
$p_x, p_y \sim |x|^{-1}$. Relevant parts of $H$ are given
below.

For the rigid-rotor part of $H$, quantization is achieved by
symmetrization with respect to $\alpha, \beta$ and by putting $p_\beta
= -i (\sin \beta)^{- 1/2} \partial_\beta (\sin \beta)^{1/2}$,
$p_\alpha = -i \partial_\alpha$. This is the usual quantization
  for a particle on the sphere. For the remaining canonical momenta
we have
\be
\label{10}
p^x_{\lambda \mu} = - i \frac{\partial}{\partial x_{\lambda \mu}} \ , \
p^y_{\lambda \mu} = - i \frac{\partial}{\partial y_{\lambda \mu}} \ .
\ee
Substitution of these expressions into Eq.~(\ref{9}) yields the
quantized form of the square $\hat{I}^2$ of the operator of total
angular momentum. The operator $\hat{K}$ is given by $\hat{K} =
\sum_{\lambda\mu} \left(x_{\lambda\mu} p^y_{\lambda\mu} -
y_{\lambda\mu} p^x_{\lambda \mu} \right)$. The three components
$\hat{I}_x$, $\hat{I}_y$, $\hat{I}_z$ of the quantized angular
momentum can be shown to obey the standard commutation
relations. Moreover, every component commutes with $\hat{K}$ and with
the quantized Hamiltonian $\hat{H}$. A complete set of commuting
operators is, therefore, $\hat{I}^2, \hat{I}_z, \hat{K}, \hat{H}$.

\subsection{Spectra} 
The leading-order (${\cal O}(\omega)$) contribution
to $\hat{H}$ is given by
\bas
\hat{H}_\omega = \sum_{\lambda \mu} \left( {(p^x_{\lambda
\mu})^2 + (p^y_{\lambda \mu})^2 \over 2C_b} + {C_b\over 2} \omega^2_{\lambda \mu}
\left( x_{\lambda \mu}^2 + y_{\lambda \mu}^2 \right) \right)
\eas
and describes a set of uncoupled harmonic oscillators with frequencies
$\omega_{\lambda \mu} = [(\lambda ( \lambda + 1) D_a + \mu^2 D_{b}) /
  C_b]^{1/2}$. We combine the degrees of freedom $x_{\lambda \mu}$ and
$y_{\lambda \mu}$ into a two-dimensional $SO(2)$ symmetric harmonic
oscillator with quantum numbers $n_{\lambda \mu} = 0, 1, 2, \ldots$
and $k_{\lambda \mu} = 0, \pm 1, \pm 2, \ldots$. In units of
$\omega_{\lambda \mu}$ the energies are $2n_{\lambda \mu} +
|k_{\lambda \mu}| +1$. The operator $\hat{K}$ has eigenvalues $K =
\sum_{\lambda\ge 2} \sum_{\mu = - \lambda}^\lambda k_{\lambda \mu}$.

Next-order corrections to the Hamiltonian $\hat{H}_\omega$ are either
of order ${\cal O}(\xi)$ or of order ${\cal O}(x^2\omega)$.
The former couple rotations to vibrations. The latter add
anharmonicities to the harmonic vibrations and thereby lift the
degeneracies. We confine ourselves here to the former. The Hamiltonian
is
\be
\label{12}
\hat{H}_{\omega,\xi}= \hat{H}_{\omega} + {\hat{I}^2 - \hat{K}^2
\over 2C_a} \ ,
\ee
with $\hat{I}^2$ given by the quantized form of Eq.~(\ref{9}). The
eigenfunctions of $(\hat{I}^2 - \hat{K}^2)$ are Wigner $D$-functions
$D_{M,K}^I(\alpha,\beta,0)$ that depend on total integer spin $I$ and
its projections $-I \le M, K \le
I$~\cite{varshalovich1988,papenbrock2011}. The eigenvalues of
$\hat{I}^2$ are $I (I + 1)$ with $I \ge |K|$. We see that each
vibrational state of the leading-order Hamiltonian $\hat{H}_\omega$
becomes a band head with spin $|K|$, and the spectrum exhibits a
rotational band on top of each band head.  At this order in the EFT,
all rotational bands have the same moment of inertia $C_a$.
Differences in the moments of inertia are higher-order
effects~\cite{Zhang2013}.

The ground state has quantum numbers $n_{\lambda \mu} = 0$,
$k_{\lambda \mu} = 0$ (this implies $K = 0$) and spin $I = 0$. It has
positive parity and, in the case of nuclei, positive ${\cal R}$
parity. For nuclei, this limits the rotational states in the
ground-state band to even values of $I$. 

We turn to excited states. Here nuclei and linear molecules differ.
For $D_b > 0$ the lowest single vibrational excitation corresponds to
the mode $(x_{2 0}, y_{2 0})$. The fields $x$ and $y$ have positive
parity, in keeping with the parity of the axial vectors $J_x$ and
$J_y$ in Eqs.~(\ref{1}). The lowest excitation has $|K| = 1$, negative
intrinsic ${\cal R}$ parity and, thus, values of $I=1,2,3,\ldots$. For
linear molecules, states with $|K| = 1$ are indeed the lowest-lying
vibrations~\cite{Herzberg1945}.  In contradistinction, in nuclei such
states are excluded because paired states have positive ${\cal R}$
parity.  Pair breaking (i.e., generation of states with odd ${\cal R}$
parity) happens only at the breakdown scale $\Lambda$ of our EFT.
Thus, pairing excludes low-lying magnetic dipole
excitations~\cite{heyde2010,bentz2011} and more generally any vibrational band
head with odd spin and positive parity in the low-energy regime.  This
essential element provides the only difference in the low-energy
spectra of molecules and of deformed nuclei. Indeed, there are no
low-lying $I^\pi = 1^+$ states in deformed even-even
nuclei~\cite{davidson1981, aprahamian2006}.

As is well known from data~\cite{davidson1981, aprahamian2006}, the
ground states of deformed even-even nuclei consistently have quantum
numbers $I = 0 = K$ and positive parity.  Low-lying vibrational states
have $K = 0$ and $|K| = 2$ and positive parity. In rare-earth nuclei,
both band heads have an excitation energy of about $1$ MeV. In the
present approach, these states are generated by local rotations around
an axis that is perpendicular to the symmetry axis of the ground
state.  However, the wavefunction corresponding to $K=0$ is symmetric
under any $SO(2)$ rotation of $x$ and $y$ and must thus be viewed as an axially
symmetric excitation that corresponds to the $\beta$ excitation of the
geometrical model~\cite{bmbook}. The $|K|=2$ wave functions exhibit no
symmetry under exchange of $x$ and $y$ and thus break the axial
symmetry. Thus, they correspond to the $\gamma$ mode of the
geometrical model.

Finally, we note that the present approach can be extended to EFTs for
other cases such as general molecules. In this case, the ground state fully
breaks $SO(3)$ to the point group ${\cal P}$ of the molecule, and the
coset space is thus $SO(3)/{\cal P}$. This case is technically simpler than 
the breaking from $SO(3)$ to $SO(2)$ because the coset essentially 
has a group structure. The coset can be parameterized by three Euler angles, 
and the zero modes indeed describe the orientation of the molecule in space 
while the Nambu-Goldstone fields describe the intrinsic vibrations.

\section{Summary} 
In summary, we have shown how to develop effective field theories for
finite systems with emergent symmetry breaking. We have applied this
approach to two types of systems. (i) Superfluids like infinite
Bose-Einstein condensates (BEC) or the paired states of a BCS
superconductor that break $U(1)$ phase symmetry. (ii) Systems with
non-spherical ground states such as molecules and atomic nuclei that
break rotational symmetry. In both cases, symmetry arguments alone
yield the universal features of the low-lying excitations. In case (i)
these are vibrations. We also relate the spectra of systems with
different particle numbers. In case (ii) these are vibrations that are
the heads of rotational bands. The moment of inertia is a fit
parameter and will, in general, be different for different physical
systems (for instance, superfluid and normal systems). Nuclei and
molecules differ in that the ground states of even-even nuclei are
paired. This accounts for the absence in nuclei of low-lying band
heads with odd spin and positive parity.

In contrast to phenomenological approaches, and except for a small
number of constants, our approach yields an explicit expression for
the Hamiltonian in leading order, and a systematic procedure to
generate terms of higher order. It may, thus, be a useful starting
point for the analysis of spectra and electromagnetic transitions.  It
is textbook knowledge that the traditional collective models of
deformed nuclei~\cite{bmbook,eisenberg,iachello} overpredict
transitions between the $\beta$-band ($\gamma$ band) and the
ground-state band by a factor of about ten (four)~\cite{rowe}.
Furthermore, the interpretation of low-lying collective $0^+$ states
as $\beta$ band heads has been put into question by the available data
on $E2$ transitions~\cite{garrett2001}. This makes it very interesting
to study electromagnetic couplings within the EFT approach.

\begin{acknowledgments}
The authors thank N. Pietralla and A. Richter for discussions.
This work has been supported by the U.S.  Department of Energy under
grant Nos. DE-FG02-96ER40963 (University of Tennessee) and
DE-AC05-00OR22725 with UT-Battelle, LLC (Oak Ridge National
Laboratory), and by the Alexander-von-Humboldt Foundation.
\end{acknowledgments}


\bibliography{ref}

\end{document}